\documentclass{elsart}
\newcommand{\be}{\begin{eqnarray}}
\newcommand{\ee}{\end{eqnarray}}
\usepackage{amsmath}
\usepackage{amssymb}
\usepackage{epsfig}

\begin{document}
\hspace{9.8 cm}FZJ--IKP(TH)--2003--17
\begin{frontmatter}
\title{How to measure the parity of the $\Theta^+$ in $\vec p\vec p$ collisions}
 
\author{C. Hanhart$^1$, M. B\"uscher$^1$, W. Eyrich$^2$, K. Kilian$^1$, U.-G.
  Mei{\ss}ner$^{1,3}$,} \author{ F. Rathmann$^1$, A. Sibirtsev$^1$, and H.
  Str\"oher$^1$}

{\small $^1$Institut f\"{u}r Kernphysik, Forschungszentrum J\"{u}lich GmbH,}\\ 
{\small D--52425 J\"{u}lich, Germany} \\

{\small $^2$ Physikalisches Institut, Universit\"at Erlangen,}\\
{\small   Erwin-Rommel-Str. 1,
D--91058 Erlangen, Gemany}\\ 

{\small $^3$Helmholtz-Institut f\"{u}r Strahlen- und Kernphysik (Theorie), 
Universit\"at Bonn}\\ 
{\small Nu{\ss}allee 14-16, D--53115 Bonn, Germany} \\ 
\begin{abstract}
  Triggered by a recent paper by Thomas, Hicks and Hosaka, we investigate
  which observables can be used to determine the parity of the $\Theta^+$ from
  the reaction $\vec p\vec p \to \Sigma^+\Theta^+$ near its production
  threshold. In particular, we show that the sign of the spin correlation
  coefficient $A_{xx}$ for small excess energies yields the negative of the
  parity of the $\Theta^+$. The argument relies solely on the Pauli principle
  and parity conservation and is therefore model--independent.
\end{abstract}

\end{frontmatter}

{\bf 1.} There is increasing experimental evidence for an exotic baryon with
strange{-}ness $S = + 1$, the $\Theta^+ (1540)$, see e.g.
\cite{LEPS,DIANA,CLAS1,Saphir,CLAS2}, preceded and complemented by a flurry of theoretical
activity, see e.g.  \cite{Walliser,DPP,Weigel,JW,lattice}. Experimental activities are now
trying to pin down the quantum numbers of the $\Theta^+$; in particular the
parity $\pi(\Theta^+)$ of this state is so far not determined experimentally, and the
theoretical predictions allow for both possibilities. For example the
chiral soliton model points at a positive parity state \cite{DPP,Weigel}
whereas  lattice calculations indicate a negative parity state for the pentaquark
ground state \cite{lattice}, just to name a few.
 It is thus of utmost
importance to determine $\pi(\Theta^+)$ to further constrain the internal
dynamics and structure of this exotic state.

The determination of the internal parity of a state is in general difficult,
for any signal gets distorted by the interference of the resonance amplitude
with the background \cite{kanzo}. Thus, in order to unambiguously pin down the
parity of a state from an angular distribution, one needs to know the
background rather precisely. This makes it difficult to get model--independent
results. A way to minimize the impact of the background would be to also
measure the polarization of the decay products of the resonance as proposed in
Ref. \cite{oset}, however, these are extremely difficult measurements. In a
recent work, Thomas, Hicks and Hosaka have proposed an alternative method to
unambiguously determine the parity of the $\Theta^+$ by looking at $\vec p\vec
p\to \Theta^+\Sigma^+$ close to the production threshold \cite{thomas}. The
idea relies only on the conservation of total angular momentum and parity in
strong interactions and is therefore completely independent of the reaction
mechanism. The same method was proposed by Pak and Rekalo to determine the
parity of the kaon \cite{rekalo2}.

In this paper we will discuss
this proposal in more detail. To be specific we will supply
first a discussion of the most relevant observable including its angular
structure and energy dependence in the near threshold regime,
and second a brief discussion  of its experimental boundary conditions as well
as its feasibility.
Let us start, however, with a repetition of the argument by Thomas et al.
It is well known that the Pauli principle closely links spin
and parity of a two nucleon state, since the relation
$(-)^{S+L+T}=-1$ holds,
where $T$ denotes the total isospin of the two nucleon system
(for $pp$ $T=1$), $L$ the angular momentum and $S$ the total spin. Thus,
a spin triplet (singlet) $pp$ pair has to be in an odd (even) parity
state. As a consequence, selecting the spin of a $pp$ state implies preparing
its parity. From the argument given it also follows that the corresponding reaction from
a $pn$ initial state does not allow to prepare the initial parity, for a $pn$
state is an admixture of $T=1$ and $T=0$ states.

Close to the production threshold only $s$--waves are kinematically allowed in
the final state. Consequently,  a negative parity $\Theta^+$ can
only originate from a spin triplet initial $pp$ state while a positive parity
$\Theta^+$ can only stem from a spin singlet $pp$ state. 
In
Refs. \cite{meyer,deepak,report} it was shown, that a measurement of 
the spin correlation parameters $A_{xx}, \ A_{yy}, \ A_{zz}$ as well
as the unpolarized  cross section allows to project on the individual initial
spin states. More precisely
\begin{eqnarray}
^1\sigma_0&=&\sigma_0(1-A_{xx}-A_{yy}-A_{zz}) \ ,\nonumber \\
^3\sigma_0&=&\sigma_0(1+A_{xx}+A_{yy}-A_{zz})
\  ,\nonumber \\
^3\sigma_1&=&\sigma_0(1+A_{zz}) \ ,
\label{spinwqs}
\end{eqnarray}
where the spin cross sections are labeled following the convention of
 Ref. \cite{meyer}
 as $^{2S+1}\sigma_{M_S}$, with $S$
the total spin of the initial state and $M_S$ its projection;   $\sigma_0$
 denotes the unpolarized cross section.
Therefore, if only $\Sigma^+\Theta^+$ $s$--waves contribute and the $\Theta^+$
has positive parity, only $^1\sigma_0$ would be non--vanishing.

Unfortunately, longitudinal polarization (needed for $A_{zz}$) is not easy to
prepare. However, the following linear combination projects on spin triplet
initial states and no longitudinal
polarization is needed:
\begin{eqnarray}
^3\sigma_\Sigma=\frac12(^3\sigma_0+^3\sigma_1)=\frac12\sigma_0(2+A_{xx}+A_{yy}) \ .
\label{dss}
\end{eqnarray}
Thus, if the $\Theta^+$ has positive parity, then 
$(A_{xx}+A_{yy})$ must go to a value of $-2$ as the energy approaches the threshold. 
Since $A_{xx}$ and $A_{yy}$ individually have to go to $-1$, in the
following   sections we will concentrate on a study of $A_{xx}$ only.

In the next section we will investigate the general structure of the amplitude
that will allow us to move away from the threshold region.  To be
specific, we will discuss in some detail the result one should get for
$A_{xx}$ in the reaction $\vec p\vec p\to \Sigma^+\Theta^+$ under the
assumption that the $\Theta^+$ has spin $1/2$. Note that, if its spin were
$3/2$, $^3\sigma_\Sigma$ would still be non--zero close to threshold only if
the parity were negative. However, the spin formalism would be more
complicated. Given the strong theoretical arguments in favor of a spin--1/2
pentaquark we do not consider this option in detail. In
addition, we take the $\Theta^+$ as being a stable state,
for the narrow width of the $\Theta^+$ will not change any of the arguments
given.

{\bf 2.} In this section we closely follow Ref. \cite{report}. (For a
discussion of the threshold region also see Ref. \cite{rekalo}.) For the general
structure
of the matrix element for $pp\to\Sigma^+\Theta^+$ we may write
\begin{equation}
{ M}=H({ I}\, { I \, '})+ i\vec Q \cdot 
(\vec S \, { I '})+i\vec A \cdot (\vec S \, '\, { I})
 +(S_i \, S_j \, ')
B_{ij} \ ,
\label{ampdef}
\end{equation}
where 
${ I}=\left(\chi_2^T\sigma_y \chi_1\right)$,
${ I \, '}=\left(\chi_4^\dagger \sigma_y (\chi_3^T)^\dagger\right)$,
 $\vec S = \chi^T_1\sigma_y\vec \sigma\chi_2$ and $\vec S \, ' =
 \chi^\dagger_3\vec \sigma \sigma_y (\chi^\dagger_4)^T$. Here
$\chi_i$ denotes the Pauli spinors for the incoming nucleons (1,2)
and outgoing baryons (3,4) and $\vec \sigma$ denotes the
usual Pauli spin matrices.
We focus on the close to threshold regime and
 restrict ourselves to a non--relativistic treatment of the
outgoing particles. This largely simplifies the formalism since
a common quantization axis can be used for the complete system.
The amplitudes $H$ and $\vec A$ correspond to a spin singlet
initial $pp$ state; $\vec Q$ and $B_{ij}$ correspond to a spin
triplet initial $pp$ state.

\begin{figure}
\begin{center}
\vskip 7cm          
\includegraphics{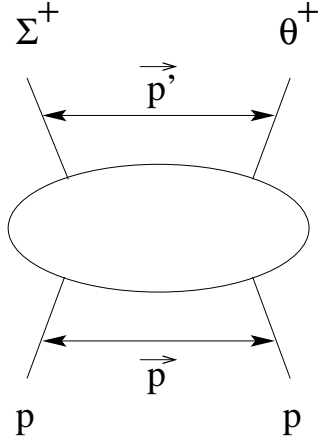} 
\caption{Illustration of the reaction.}
\label{ill}
\end{center}
\end{figure}

The whole system is characterized by the initial (final) cms momentum vector
$\vec p$ ($\vec p\, '$) as well as the axial vectors $\vec S$ and $\vec S\,
'$.  In order to construct the most general transition amplitude that
satisfies parity conservation, we combine the vectors and axial
vectors given above such that the final expressions form a scalar (pseudoscalar)
for reactions where the final state has positive (negative) intrinsic parity.
 As described in the introduction, the Pauli
principle requires $H$ and $\vec A$ ($\vec Q$ and $B_{ij}$) to be even (odd) in
$\vec p$. These two conditions together strongly constrain the number of allowed
structures for the various amplitudes. To make the notation transparent we
will use the parity of the $\Theta^+$ as a superscript to the amplitudes. 
Keeping only terms which are at most quadratic in $\vec p\, '$
we write for
 a positive parity $\Theta^+$
\begin{eqnarray}
\nonumber
H^+&=& h_0^+ \ , \\
\nonumber
\vec A^+&=& \frac{a_0^+}{\Lambda^2}(\hat p\times \vec p\, ')(\hat p\cdot \vec p\, ') \ , \\
\nonumber
\vec Q^+&=& \frac{q_0^+}{\Lambda}(\hat p\times \vec p\, ') \ , \\
B_{ij}^+&=&\frac1{\Lambda}\left(b_0^+\delta_{ij}(\hat p\cdot \vec p\, ')+
b_1^+p_ip'_j+b_2^+p_jp'_i+b_3^+p_ip_j(\hat p\cdot \vec p\, ')\right) \ ,
\end{eqnarray}
 and in case of a negative parity $\Theta^+$
\begin{eqnarray}
\nonumber
H^-&=& 0 \ , \\
\nonumber
\vec A^-&=& \frac1{\Lambda}\left(a_0^-\vec p\, '+
a_1^-\hat p(\hat p\cdot \vec p\, ')\right) \ , \\
\nonumber
\vec Q^-&=& q_0^-\hat p+\frac{q_1^-}{\Lambda^2}\vec p\, '
(\hat p\cdot \vec p\, ') \ ,  \\
B_{ij}^-&=&\epsilon^{ijk}
\left(b_0^-\hat p_k+\frac{b_1^-}{\Lambda^2}\vec p_k\, '(\hat p\cdot \vec p\, ')\right)
+\frac1{\Lambda^2}(\hat p\times \vec p\, ')_j(b_2^-\vec p_i\, '+b_3^-\hat
p_i(\hat p\cdot \vec p\, ')) \ ,
\end{eqnarray}
where $\hat p=\vec p/|\vec p|$ and the scale $\Lambda$ was introduced to make
the dimensions of the coefficients equal.  Below we estimate the size of
$\Lambda$ on dimensional grounds. Then one expects all amplitudes to be of the
same order of magnitude.  The coefficients are functions of $p^2$, $p'\, ^2$
and $(\vec p\cdot \vec p\, ')^2$.  To eliminate linearly dependent structures
the reduction formula given in Appendix B of Ref. \cite{report} may be used.

In the near threshold regime the only amplitudes that contribute
 for a negative parity $\Theta^+$ are
 $q_0^-$ and $b_0^-$, corresponding to the transition $^3P_0\to^1S_0$
and $^3P_1\to^3S_1$, respectively. Analogously, for a
positive parity $\Theta^+$ the only contributing amplitude is $h_0^+$, corresponding to
$^1S_0\to^1S_0$. 

For our discussion we express $A_{xx}$ in terms of the amplitudes
given above:
\begin{eqnarray}
4\sigma_0 &=&\phantom{+} |H|^2+|\vec Q|^2+|\vec A|^2+|B_{mn}|^2 \, , \label{si0def}
 \\
4A_{xx}\sigma_0 &=& 
-|H|^2-2|Q_x|^2+|\vec Q|^2-|\vec A|^2-2|B_{xm}|^2+|B_{nm}|^2 \ .
\label{axx}
\end{eqnarray}
where summation over $m$ and $n$ is assumed.
Keeping only terms up--to--and--including $p'\, ^2$
we
obtain for a positive parity $\Theta^+$
\begin{equation}
A_{xx} = -1 + \alpha \, \frac{p'\, ^2}{\Lambda^2} + O\left(\frac{p'\, ^4}{\Lambda^2}\right)
\end{equation}
where
\begin{eqnarray}
\nonumber
\alpha &=& \frac2{|h_0^+|^2}\left(
|q_0^+|^2\sin^2(\theta)\cos^2(\phi)+
|a_0^+|^2\sin^2(\theta)\cos^2(\theta)+\right.\\
\nonumber
& &\phantom{\frac2{|h_0^+|^2}(}
\left.(|b_1^+|^2+|b_2^+|^2\sin^2(\phi))\sin^2(\theta)+\right.\\
& &\phantom{\frac2{|h_0^+|^2}(}\left.
(|b_0^+|^2+|b_0^++b_1^++b_2^++b_3^+|^2)\cos^2(\theta)
\right) \ .
\label{alpha}
\end{eqnarray}
This expression reproduces the threshold behavior of $A_{xx}$
(approaching $-1$) as discussed in the first section.
Exactly at threshold the cross
section vanishes, however,  Eq. (\ref{alpha}) allows one to estimate the energy
dependence
of $A_{xx}$ based on rather general arguments.

In case of a negative parity $\Theta^+$ we have
\begin{equation}
A_{xx} = \frac{|q_0^-|^2}{|q_0^-|^2+2|b_0^-|^2}\left(1 -
 \beta \, \frac{p'\, ^2}{\Lambda^2} + O\left(\frac{p'\, ^4}{\Lambda^2}\right)\right)
\end{equation}
where
\begin{eqnarray}
\nonumber
\beta &=& \frac1{|q_0^-|^2}\left(
-2\mbox{Re}(q_0^{-\, *}q_1^-)\cos^2(\theta)+
(|a_0^-|^2\sin^2(\theta)+|a_0^-+a_1^-|^2\cos^2(\theta))\right.\\
\nonumber
& &\phantom{\frac2{|h_0^+|^2}(}\left.
+2\mbox{Re}(b_0^{-\, *}b_2^-)\sin^2(\theta)(\cos^2(\phi)-\sin^2(\phi))\right)\\
\nonumber
&+& \frac1{|q_0^-|^2+2|b_0^-|^2}\left(
2\mbox{Re}(q_0^{-\, *}q_1^-+2b_0^{-\, *}b_1^-)\cos^2(\theta)\right.\\
\nonumber
& &\phantom{\frac2{|h_0^+|^2}(}\left.
+(|a_0^-|^2\sin^2(\theta)+|a_0^-+a_1^-|^2\cos^2(\theta))+\right.\\
& &\phantom{\frac2{|h_0^+|^2}(}\left.
+2\mbox{Re}(b_0^{-\, *}b_2^-)\sin^2(\theta)\right) \ .
\label{beta}
\end{eqnarray}
This formula only holds for non--vanishing values of $q_0^-$. (If
$q_0^-$ would vanish, the denominator in the third row of
this expression needs to be replaced by $1/|q_0^-|^2$.)

Because of the presence of two amplitudes even at threshold the assymptotic
value
of $A_{xx}$ is not equal to $+1$. In this case $A_{xx}$
measures the ratio of the two amplitudes.  

Let us briefly discuss the angular dependence of $A_{xx}$.
We observe that for both scenarios there is no preferred angular
range. Thus one can perform the experiment angular integrated or only for special
kinematics of the $\Sigma^+\Theta^+$ final state. The information content will be the same.
Secondly, one might ask wether it is sufficient to measure $A_{xx}$ at some single
low excess energy and then verify that only $s$--waves contribute by analysing
 the angular distribution. The above formulas show that
this will not work. If only the $p$--wave amplitudes $b_1^+$ ($a_0^-$)
contribute considerably in case of a positive (negative) parity pentaquark,
the differential
cross section would still  be isotropic. It is therefore nescessary to study
the energy dependence of $A_{xx}$, as was also stressed in Ref. \cite{thomas}.

{\bf 3.} The next step is to estimate the energy range where the formulas
given in the previous section should give sensible results. 

\begin{figure}
\begin{center}
\vskip 4cm          
\includegraphics{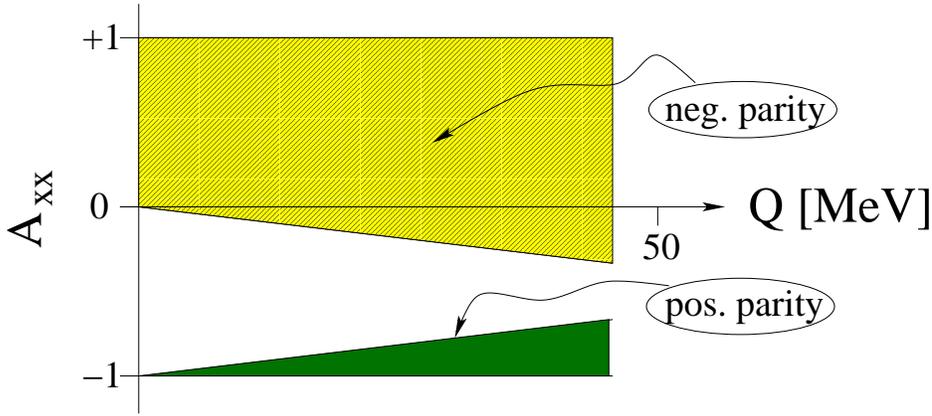} 
\caption{Schematic presentation of the
 result for $A_{xx}$  for the two possible parity states of the $\Theta^+$.
For either option realized the corresponding data should fall into the area indicated.
  In case of a negative parity the threshold value depends on the ratio of the
  strength of the two possible $s$--wave amplitudes.}
\label{sketch}
\end{center}
\end{figure}

The reaction $\vec p\vec p \to \Sigma^+\Theta^+$ near threshold is
characterized by a large momentum transfer $t\sim -m_xM_N$, where $m_x$
denotes the mass produced $m_x=M_\Sigma+M_\Theta-2M_p$ and $M_N \,(M_\Sigma)$
denotes the nucleon ($\Sigma$) mass.  Since in a non--relativistic picture $t$
is a measure of the size of the interaction region, it sets the scale for the
onset of higher partial waves. Thus, we can identify $\Lambda^2$ with $-t$. We
express $p'$ in terms of $Q$, the excess energy above the $\Sigma^+\Theta^+$
threshold, $Q=p'\, ^2/(2\mu)$, where $\mu$ denotes the reduced mass of the
$\Sigma^+\Theta^+$ system, and find that
$$
\frac{p'\, ^2}{\Lambda^2}\sim 2 \, \frac Q{M_N} 
$$
For our estimates of the energy
dependence of the polarization observables we need $(p'/\Lambda)^2\ll 1$.
If we request $(p'/\Lambda)^2\sim 0.1$ we find that the expressions given above
should be applicable for $Q<50$~MeV.
The expected signal for $A_{xx}$ is sketched in Fig. \ref{sketch}.

Implicitly we assumed that there is no strong $\Theta^+\Sigma^+$ final state
interaction that would introduce an additional large scale into the system.
This is justified, because most of those meson exchanges that potentially
could lead to a strong final--state interaction are either absent or should be
weak: (i) a single pion exchange between $\Theta^+$ and $\Sigma^+$ is not
possible due to the isoscalar nature of the pentaquark, (ii) a strong coupling
of the $\Theta^+$ to $NK$ is excluded due to its small width and (iii) there
can also be no strong coupling of the $\Theta^+$ to the iso--scalar two pion
exchange, known to be responsible for the medium range attraction of the $NN$
interaction, since then the $\Theta^+$ should not be seen equally narrow in
nuclear reactions \cite{DIANA} and in elementary production reactions on a
single nucleon \cite{Saphir,CLAS2}.

{\bf 4.} We also want to briefly comment on the possible influence of the
background on the signal. In principle there is the admittedly rather unlikely
 possibility that the
pentaquark signal observed does---at least to some extend---stem from an
interference of the $\Theta^+$ production amplitude with the background. How
would this change our analysis? To simplify this discussion we will assume
that the observables are fully angular integrated. Thus,  we do not have to worry
about the interference amongst different partial waves. In addition, we will
only discuss the observables in threshold kinematics.
 
 The observable $^3\sigma_\Sigma$ defined in
Eq. (\ref{dss}) is a projector on spin triplet initial states irrespective of
the final states. Thus, even if the observed signal would be due to an
interference of a positive parity $\Theta^+$ amplitude with the background,
$A_{xx}$ would approach $-1$ as the energy approaches the $\Theta^+$
production threshold. 

The situation is a little more complicated for the negative parity $\Theta^+$,
for here two amplitudes contribute at threshold. In this case the threshold
amplitude would read
\begin{equation}
A_{xx} = \frac{2\mbox{Re}(q_B^{-\, *}q_R^-)+|q_R^-|^2}
{2\mbox{Re}(q_B^{-\, *}q_R^-+2b_B^{-\, *}b_R^-)+
|q_R^-|^2+2|b_R^-|^2} \, ,
\label{interfer}
\end{equation}
where we introduced the following decomposition (analogously for $b_0^-$)
$q_0^-=q_B^{-}+q_R^{-}$, where the first and second term denote the
background and the resonance amplitude, respectively. In addition we assumed that
all those terms that are at least linear in the resonance amplitude are
(miss)interpreted as $\Theta^+$ resonance signal. The sign of $\mbox{Re}(q_B^{-\, *}q_R^-)$ is
not fixed.  Thus, if the interference term dominates over the pure resonance
contribution, the value of $A_{xx}$ can be negative even if the pentaquark has
negative parity. However, the denominator of $A_{xx}$ in Eq.  (\ref{interfer})
denotes the differential cross section and is thus bound to be positive
(otherwise, already the unpolarized measurement would tell that the signal is
dominated by an interference) and consequently the term $(2\mbox{Re}(b_B^{-\,
  *}b_R^-)+|b_R^-|^2)$ must be positive. In this case it is
straightforward to show that the depolarization coefficient $D_{zz}$ as well
as the spin correlation coefficient $A_{zz}$ both take unphysical values
larger than 1.  On the other hand one finds for these observables in case of a
positive parity $\Theta^+$ the asymptotic values 0 and -1 for $D_{zz}$ and
$A_{zz}$, respectively.

Thus, a simultanous investigation of $A_{xx}$ and either $D_{zz}$ or $A_{zz}$
allows to unambigously determine the parity of the $\Theta^+$ even for the
unlikely situation that an interference term is missinterpreted as a resonance
signal.

A similar discussion for the possible influence of background amplitudes on
the interpretation of polarization observables in terms of the $\Theta^+$
parity for other reactions is urgently called for. Within particular models
this was done in Refs. \cite{oset,kanzo}.

{\bf 5.} We would like to make some comments about the possible
experimental realization. Double polarization experiments with polarized proton beams with
momenta up to 3.7 GeV/c and polarized internal or external targets can
be carried out at the COoler SYnchrotron COSY-J\"ulich. Two
experimental facilities, the magnetic spectrometer ANKE and the
time-of-flight spectrometer TOF, can be used for such
measurements. Since both cannot detect photons the relevant
reaction channels are: $$\Theta^+\Sigma^+\to
\Sigma^+\left[pK^0\right]\to \left({p\pi^0 \atop n\pi^+}\right)\left[p(\pi^+\pi^-)\right]\ .$$
This implies  $K^0$ identification by the invariant mass and the $\Sigma^+$ by
missing mass and asking for an additional proton ($\Sigma^+\to p\pi^0$) or an
additional positive pion ($\Sigma^+\to n\pi^+$).
For the candidate events $\{\Sigma^+,K^0\}$ the invariant mass of the $[pK^0]$
subsystem has to be reconstructed.

The $\Theta^+$ production cross section from the reaction
$pp{\to}\Sigma^+\Theta^+$ has been estimated within a meson exchange model
calculation~\cite{Polyakov} taking the $\Theta^+$ width of 5~MeV. Its value at
$Q{\le}$100~MeV above the $\Sigma^+\Theta^+$ threshold is about 80-120 nb,
which is about a factor of 20 smaller than the most recent
estimation~\cite{Liu} using a $\Theta^+$ width of 20~MeV.  At TOF a 5$\sigma$
signal in the $K^0p$ invariant mass distribution has been observed in the
reaction $pp\to \Sigma^+K^0p$ at a beam momentum of 2.95 GeV/c, corresponding
to an excess energy well below 50 MeV. The very preliminary cross section
estimate is of the order of a few hundred nb.  The width of the peak is about
25 MeV, corresponding to the experimental resolution of TOF. At ANKE a
proposal has been accepted to measure the reaction $pp\to K^0 p \Sigma^+$
\cite{anke}. These measurements will be carried out in spring 2004.

For the envisaged future double polarization experiments at COSY a frozen spin
NH$_3$ target (TOF))\cite{fst} and a polarized internal hydrogen gas target
utilizing a storage cell (ANKE) \cite{abs} are presently being developed.

{\bf 6.} To summarize, we have extended
the argument of Ref.~\cite{thomas} and specified
how the parity of the $\Theta^+$ can be determined in the reaction $\vec p\vec
p\to \Sigma^+ \Theta^+$. In particular, we have identified an ideally suited
observable, namely the spin correlation coefficient $A_{xx}$,
whose sign agrees with $\pi (\Theta^+)$ near threshold. We have
further discussed the information content of the angular distribution
of $A_{xx}$, and
identified the relevant energy range for the measurement. If the final--state
interaction between the $\Sigma^+$ and the $\Theta^+$ is weak, as can be
assumed given the known or anticipated properties of the $\Theta^+$, the
above mentioned identification of the parity with the sign of $A_{xx}$ holds
for an excess energy (with respect to the $\Sigma^+\Theta^+$ threshold) below
50 MeV. Finally, we have briefly discussed the experimental feasibility of
this proposal and shown that such a determination of the 
$\Theta^+$ parity
$\pi (\Theta^+)$ is
indeed possible at COSY.

{\bf Note added:} After the original submission of this letter a model calculation
for $A_{xx}$ for the reaction $\vec p\vec p\to \Sigma^+ \Theta^+$
appeared \cite{axxmodel}. The results of this phenomenological approach lie well within the
range given in Fig. \ref{sketch}. 

{\bf Acknowledgments}

We thank M. Polyakov for useful discussions about the possible role of the
$\Sigma^+\Theta^+$ final state interaction.


\begin{thebibliography}{10}

\bibitem{LEPS}
T.~Nakano {\it et al.}  [LEPS Collaboration],
Phys.\ Rev.\ Lett.\  {\bf 91} (2003) 012002
[arXiv:hep-ex/0301020].
\bibitem{DIANA}
V.~V.~Barmin {\it et al.}  [DIANA Collaboration],
Phys.\ Atom.\ Nucl.\  {\bf 66} (2003) 1715
[Yad.\ Fiz.\  {\bf 66} (2003) 1763]
[arXiv:hep-ex/0304040].
\bibitem{CLAS1}
S.~Stepanyan {\it et al.}  [CLAS Collaboration],
arXiv:hep-ex/0307018.
\bibitem{Saphir}
J.~Barth {\it et al.}  [SAPHIR Collaboration],
Phys.\ Lett.\ B {\bf 572} (2003) 127.
\bibitem{CLAS2}
V.~Kubarovsky {\it et al.}  [CLAS Collaboration],
arXiv:hep-ex/0311046.
\bibitem{Walliser}
H.~Walliser,
Nucl.\ Phys.\ A {\bf 548} (1992) 649.
\bibitem{DPP}
D.~Diakonov, V.~Petrov and M.~V.~Polyakov,
Z.\ Phys.\ A {\bf 359} (1997) 305
[arXiv:hep-ph/9703373].
\bibitem{Weigel}
H.~Weigel,
Eur.\ Phys.\ J.\ A {\bf 2} (1998) 391
[arXiv:hep-ph/9804260].
\bibitem{JW}
R.~L.~Jaffe and F.~Wilczek,
Phys.\ Rev.\ Lett.\  {\bf 91} (2003) 232003
[arXiv:hep-ph/0307341].
\bibitem{lattice}
F.~Csikor, Z.~Fodor, S.~D.~Katz and T.~G.~Kovacs,
JHEP {\bf 0311} (2003) 070;
S.~Sasaki,
arXiv:hep-lat/0310014.
\bibitem{kanzo}
K.~Nakayama and K.~Tsushima,
arXiv:hep-ph/0311112.
\bibitem{oset}
T.~Hyodo, A.~Hosaka and E.~Oset,
arXiv:nucl-th/0307105.
\bibitem{thomas}
A.~W.~Thomas, K.~Hicks and A.~Hosaka,
arXiv:hep-ph/0312083.
\bibitem{rekalo2}
N.~K.~Pak and M.~P.~Rekalo,
Phys.\ Rev.\ D {\bf 59} (1999) 077501.
\bibitem{meyer}
H.~O.~Meyer {\it et al.},
Phys.\ Rev.\ C {\bf 63} (2001) 064002.
%
\bibitem{deepak}
P.~N.~Deepak and G.~Ramachandran,
Phys.\ Rev.\ C {\bf 65}(2002) 027601 .
\bibitem{report}
C.~Hanhart,
arXiv:hep-ph/0311341.
\bibitem{rekalo}
E.~Tomasi-Gustafsson and M.~P.~Rekalo,
Phys.\ Part.\ Nucl.\  {\bf 33} (2002) 220
[Fiz.\ Elem.\ Chast.\ Atom.\ Yadra {\bf 33} (2002) 436].
\bibitem{Polyakov}
         M.V. Polyakov et al., Eur. Phys. J. A{\bf 9} (2000) 115.
\bibitem{Liu}
         W. Liu and C.M. Ko, Phys. Rev. C {\bf 68} (2003) 045203.
\bibitem{anke}
V.Koptev et al., COSY proposal No. 130
(2003), available via
http://www.fz-juelich.de/ikp/anke/doc/doc/Proposals.shtml
\bibitem{fst} M.
Pl\"uckthum et al., NIM A {\bf 400} (1997) 133; M. Pl\"uckthum, PhD thesis Bonn,
1998.  
\bibitem{abs} F. Rathmann et al., Proc. of the 15$\rm ^{th}$ Int. Spin
Physics Symp., Upton, New York 2002, AIP Conf. Proc. {\bf 675}, p. 924, eds.
Y.I. Makdisi, A.U. Luccio, W.W. Mackay, New York (2003).
\bibitem{axxmodel}
S.~I.~Nam, A.~Hosaka and H.~C.~Kim,
arXiv:hep-ph/0401074.
\end{thebibliography}
\end{document}